# On the space and time evolution of regular or irregular human heart or brain signals

Çağlar Tuncay
Department of Physics, Middle East Technical University
06531 Ankara, Turkey
caglart@metu.edu.tr

**Abstract:** A coupled map is suggested to investigate various spatial or temporal designs in biology: Several cells (or tissues) in an organ are considered as connected to each other in terms of some molecular diffusions or electrical potential differences and so on. The biological systems (groups of cells) start from various initial conditions for spatial designs (or initial signals for temporal designs) and they evolve in time in terms of the mentioned interactions (connections) besides some individual feedings.

The basic aim of the present contribution is to mimic various empirical data for the heart (in normal, quasi-stable, unstable and post operative physiological conditions) or brain (regular or irregular; for epilepsy) signals. The mentioned empirical data are borrowed from various literatures which are cited. The suggested model (to be used besides or instead of the artificial network models) involves simple mathematics and the related software is easy. The results may be considered as in good agreement with the mentioned real signals.

**Introduction:** It is known that the shapes (patterns) of biological units and the electrical signals coming out of them are important and extensively studied in various fields of science. (Reference [1] is a limited selection for a physicist.) Secondly, the irregular designs may be carrying some signatures of several illnesses [1-3]. It is clear that understanding the underlying mechanisms for the formation of both, the regular or irregular spatial or temporal patterns is of vital importance.

We suggest a model where various designs are obtained theoretically and they mimic various empirical spatial [4] or temporal formations in biology: A group of biological units (cells or tissues) are represented by the entries of a lattice where the (mutual) interactions are taken into account in terms of the connections (diffusions of nutrients, morphogens, and ingredients) between the entries of the lattice and each unit may also be effected individually at a time, in terms of the activators or inhibitors (catalyzes) or morphogenesis and so on. Assuming several processes for the time evolution of the patterns; several parameters for the growth of the given group of biological units are computed in terms of iterations. Secondly, the bio-electrical signals such as the electrocardiograms (ECG), electroencephalograms (EEG), etc., may be handled within the same approach. The theoretical results may be helpful for investigating some possible clues for the causal mechanisms shaping the regular or irregular patterns and signals. The model is simple and it involves no differential equation; the related mathematics is algebra and the given algorithm for the computations (or simulations) is easy. The model is presented in the following section and the applications with their results are given in the next one. The last section is devoted for discussion and conclusion.

**Model:** The biological individuals may be considered as the entries (I,J) of a square (NxN) lattice C(I,J;T) (with $I \le N$, $J \le N$) at a time T. It is known that the mentioned units are similar to the neighbor ones. Thus, many (if not all) biological aspects of each unit (cell, tissue in an organ, etc.) may be approximated as an average of these of the neighbors within various ranges. In the iterative interaction tours (time, T) the nearest neighbor (nn) entries (K,L) interact with the entry (I,J) and the parameters for C(I,J;T) average in the mean time. (For a review about various averaging processes please see [5]) Furthermore, some feeding terms which represent the growth (or decay) of the units, may be changing in time and it may be taken as variable; F(I,J;T). Hence, the biological mood (state) of the unit at (time or) the biological stage T (in years, months, days, seconds, etc.) may be described in terms of the mentioned mood at (T-1) and the sum of the effects upon it (interactions c(I,J;K,L) and the feeding terms F(I,J;T-1)) as;

$$C(I,J;T) = F(I,J;T-1) + (C(I,J;T-1) + \sum_{K,L}^{nn} c(I,J;K,L)C(K,L;T-1))/(\rho+1) \quad , \qquad (1)$$

where ρ is the number of the nn for (I,J) and the sum is over the mentioned nn entries. The range of the interactions and thus the number of the nn entries (ρ) in Eq. 1 may be selected arbitrarily. Please note that the Eq. (1) is a “coupled map” for the real C(I,J;T); while for integer (in particular, binary) it would be called "cellular automata" ([4] and the references given therein).

*Generation of signal designs*: Each entry of the lattice C(I,J,T) may be taken not only as a part of a spatial pattern (such as the formations on skin and so on [4]) but also as a biological unit producing some electromotive force or voltage difference (V) with respect to a ground value. In other words, the mentioned electrical properties may be product of the same (or similar) processes which govern the spatial formations. Thus, we may take each column (or row) of the lattice C(I,J;T) as a time series for a pulse $V_I(t;T)$ (or $V_J(t;T)$) in an electrical signal produced by the biological units (channels) sited at I=1, 2, ..., N (and similarly for J) where t (J→t or I→t) is the time for the signal which emerges at the biological stage T. In this manner, the numerical values of the entries within each column (I) or row (J) of the lattice may be considered as a part of an output (beat) for the electrical signal with duration (time period) equals to N in arbitrary unit. Then we have V(t;T)=C(I,J;T) for the pulses coming from different emitters (channels) where t is an integer variable which increases by one:

$$t=(I-1)N + J \quad . \qquad (2)$$

In the Eq. (2), I and J increase in a nested manner; that is, J increases from 1 to N for I=1 and then I increases to 2 and J increases from 1 to N again for this new value for I. This goes on till I=N, J=N; thus t (Eq. (2)) increases by one from 1 up to $N^2$; i.e., t=1,2, …, $N^2$.

The initial conditions for C(I,J;T=0) may be similarly treated as the initial signals V(t;T=0) = C(I,J;T=0) where t is the same as defined in the Eq. (2).

**3. Application and results:** The model may be used with several parameters, initial or boundary conditions and so on for mimicking the biological (or non biological) spatial or temporal designs [4]. Here, a square (30x30) lattice (C(I,J;T)) with I≤30 and J≤30 is taken to represent a group of biological units at a temporal stage designated by T where T is an integer variable which increases by unity, i.e., T=1,2, … . Secondly, some sinusoidal initial conditions (initial signals) are assumed:

$$C(I,J;T=0)=\sin(2\pi J/30) \text{ or } C(I,J;T=0)=\sin(2\pi I/30) \qquad (3)$$

for each I or J, respectively. Hence, the initial signal becomes V(t;0)=Asin(2πt/P) where P is the period of the sinusoidal signal with P=N=30 and t is an integer variable which increases by unity (Eq. (2)). Uniform coupling (with the relative strength equals to 1/2) and periodic boundary conditions are assumed within the first nn approximation, where the first nn entries of an entry sited on a side (row wise or column wise) of the lattice, involve the nn entries from the opposite side (row wise or column wise) and ρ=4 in Eq. (1). The feeding terms for the individuals (F(I,J;T)) may be positive for the activators (or negative for the inhibitors, which are excluded here) with some non random or random terms. Here, we assume time independent feedings for both of the cases considered in this section;

$$F(I,J;T)= B + R\lambda \quad . \tag{4}$$

In the Eq. (4), B and R are some (positive) real numbers and $\lambda$ is a uniform real random number with $0\leq\lambda<1$. The parameters (B, R and T) for the following applications are given in the Table I. The other model parameters are the same as mentioned before.

Note that the given parameters are not unique and various different sets of parameters may be used for similar outcomes. For example, for bigger feedings or connections the evolutions speed up and the formations ocuurs in a smaller period of time (T), or vice versa. The random terms are nedeed since the heart or brain signals involve randomnes as further discussed in the following sections.

**Case 1 Regular or irregular (for an illness) heart beats:** Figure 1 exhibits the empirical data for human heart signals, which are borrowed from [2]. The pulse waveforms are recorded for the following physiological conditions, where the time series correspond to normal (a), quasi-stable (b), unstable (c) and post-operative (stable) (d) cases. One may notice by inspection of the Figures 1 (a)-(d) that the plots are composed of somewhat similar (repeated) pulses which are more or less periodic, except the plot in the Fig. 1 (c) which is nearly a random signal. The Figures 2 (a)-(d) are the model heart signals with the initial signal defined in Eq. (3) in all and the parameters given in the Table I, where no random component is involved (R=0) for the Figs. 2 (a) and (d). Hence, the design of each pulse within the plots of the Figs. 2 (a) and (d), and these of the whole signals may be computed (not simulated) in terms of few sums of sinus terms (Eq. 3) and some constants (Eq. 4) with T=2 and T=11 in (Eq. 1), respectively where t is as defined in the Eq. 2. The figures 2 (b)-(c) are simulations, where R is the biggest for the Fig. 2 (c).

**Case 2 Regular or irregular (epileptic) brain signals:** The Figure 3 (a) shows some theoretical results with B=0.0 and R=1.0 where the initial wave (Eq. (3)) is taken as $V(t;0) = (C(I,J;T=0)=) - \sin(2\pi J/30)$. The number of iterative tours T for the signals in the Fig. 3 (a) starts with T=1 and increases by 10 up to 91 from the bottom to the top. In the mean time, the pulses become more similar to each other as T increases. The Figure 3 (b) shows the model brain signals $V_I(t;11)$ which may be considered as mimicking the real signals for epilepsy (Please see the related figures in [3]) and the mentioned pulses or the signals may be computed in terms of 11 secular equations defined by the Eq. (1) with R=0 and T=11 (Table I).

**Discussion and conclusion**:

The model is applied to several cases for the spatial or temporal formations [4]; specifically the solar darkening of human skin or growth of crystals and so on, which are not shown here. The basic aim of the present contribution is to mimic various empirical data for human heart or brain pulses The theoretical results (Figs. 2 (a)-(d) or Figs. 4 (a) and (b)) may be considered as in good agreement with the real ones (Figs. 1 (a)-(d) or Figs. 3 (a) and (b), all respectively). Yet, inspection of the figures by eyes clearly may not be a proper method to compare the plots. Some statistical data for the distribution of the amplitudes of the mentioned model signals are given as the insets within the related figures. It is known that neurons and cardiac muscle (besides skeletal muscle) are similar in many respects ([4] and the references given therein).

## TABLES

| Explanation | B | R | T |
|---|---|---|---|
| Fig. 2 (a); heart | 5 | 0 | 2 |
| Fig. 2 (b); heart | 5 | 0.5 | 31 |
| Fig. 2 (c); heart | 3.5 | 2 | 11 |
| Fig. 2 (d); heart | 5 | 0 | 11 |
| Fig. 3 (a); brain | 0 | 1 | various |
| Fig. 3 (b); brain (epilepsy) | 3.5 | 0 | 1 |

**Table 1** The parameters for the model heart or brain are given column wise in the same but in arbitrary units, where B and R are relative to the amplitude of the sinus wave in the Eq. (3) which is unity.

## FIGURES

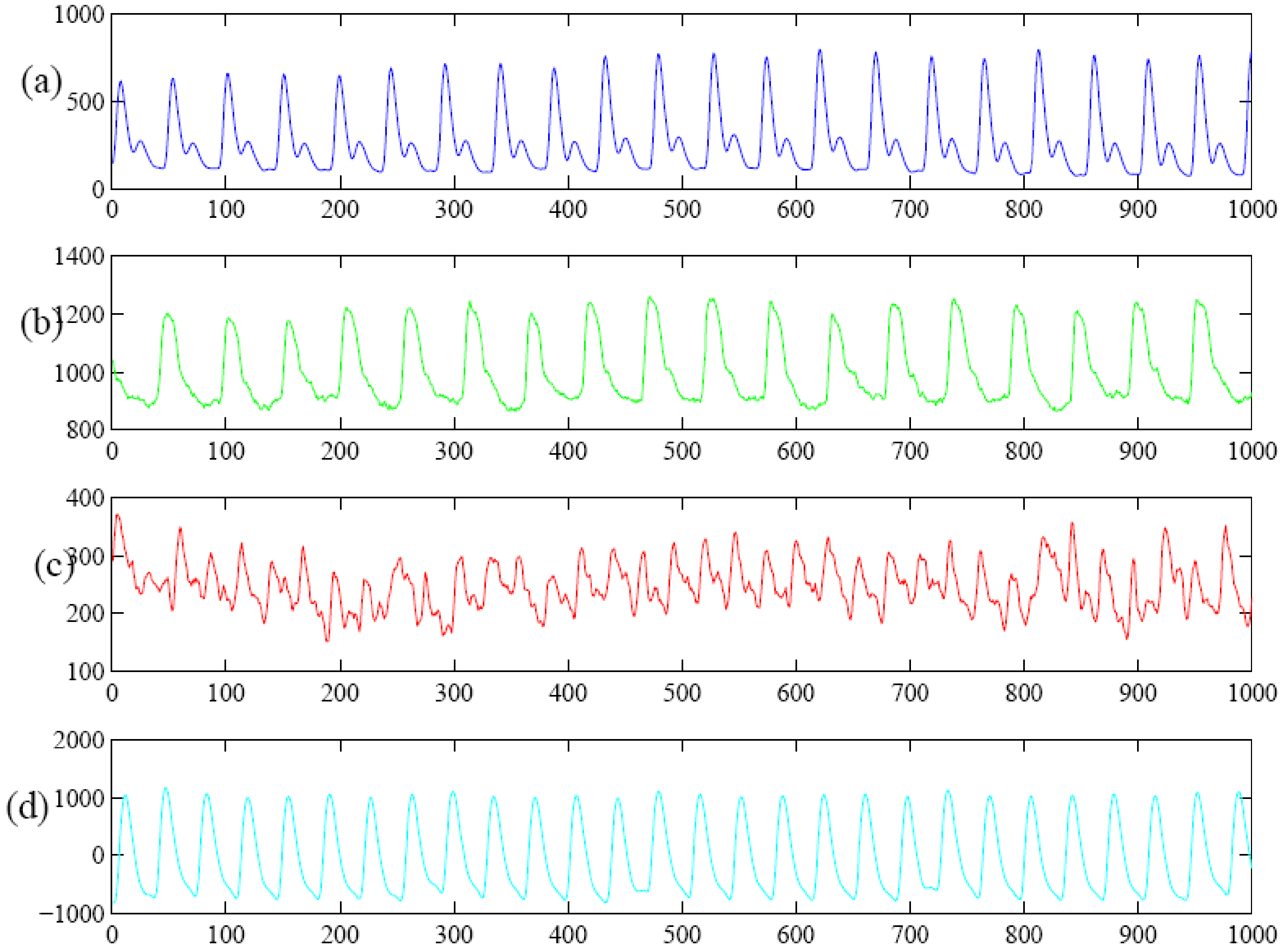

**Figure 1** The empirical data for human pulse waveform recorded with photo-plethysmography for the following physiological conditions: (a) normal, (b) quasi-stable, (c) unstable, and (d) post-operative (stable). (Borrowed from [2].)

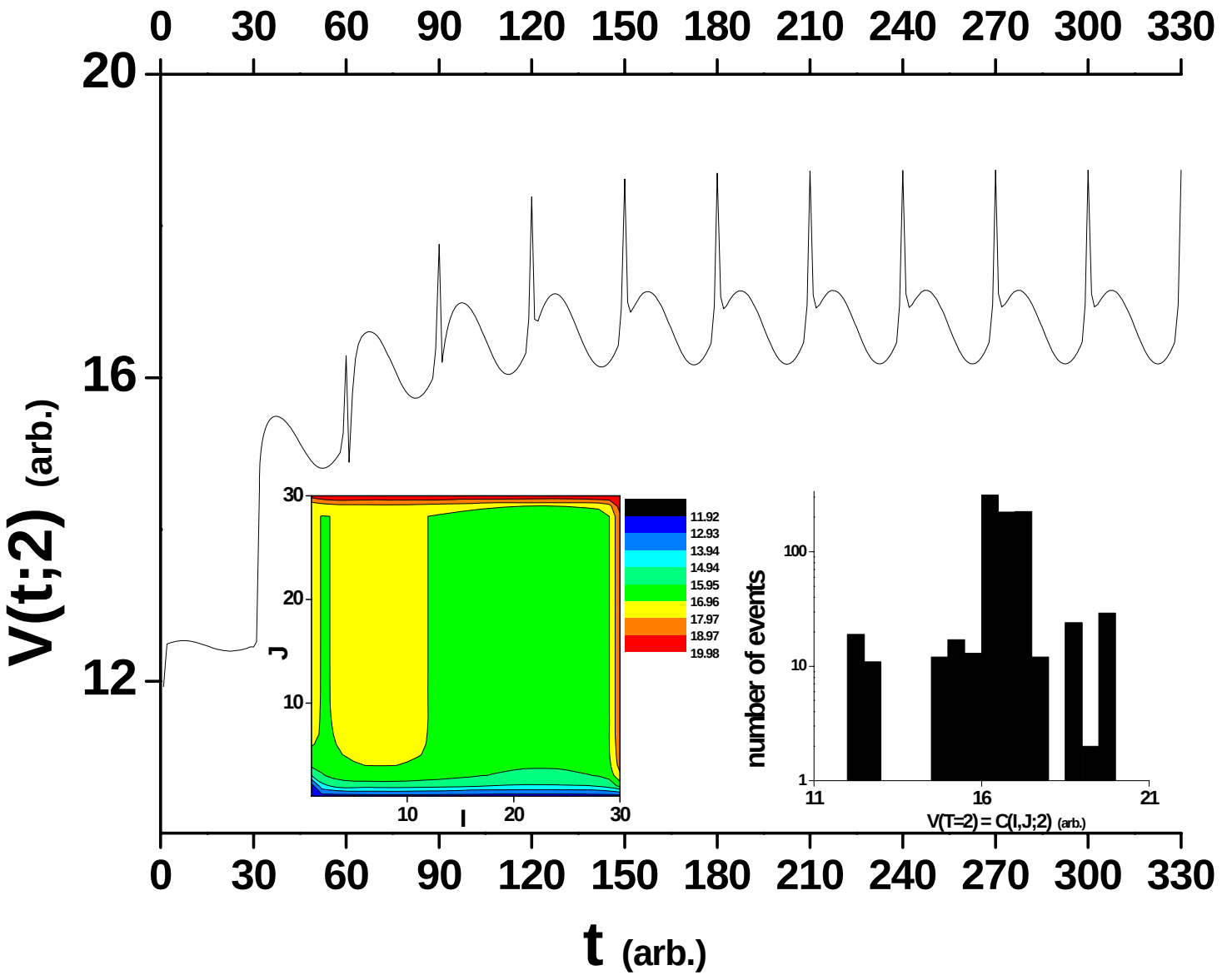

**Figure 2 (a)** Various model heart beats with the parameters given in the Table I, where there is no random parameter and the design of each pulse and this of the whole signal may be computed in terms of two (since T=2) secular equations (Eq. 1). The insets at the left and right show C(I,J,2) and the distribution of the amplitudes C(I,J,2) in semi logarithmic scale, respectively.

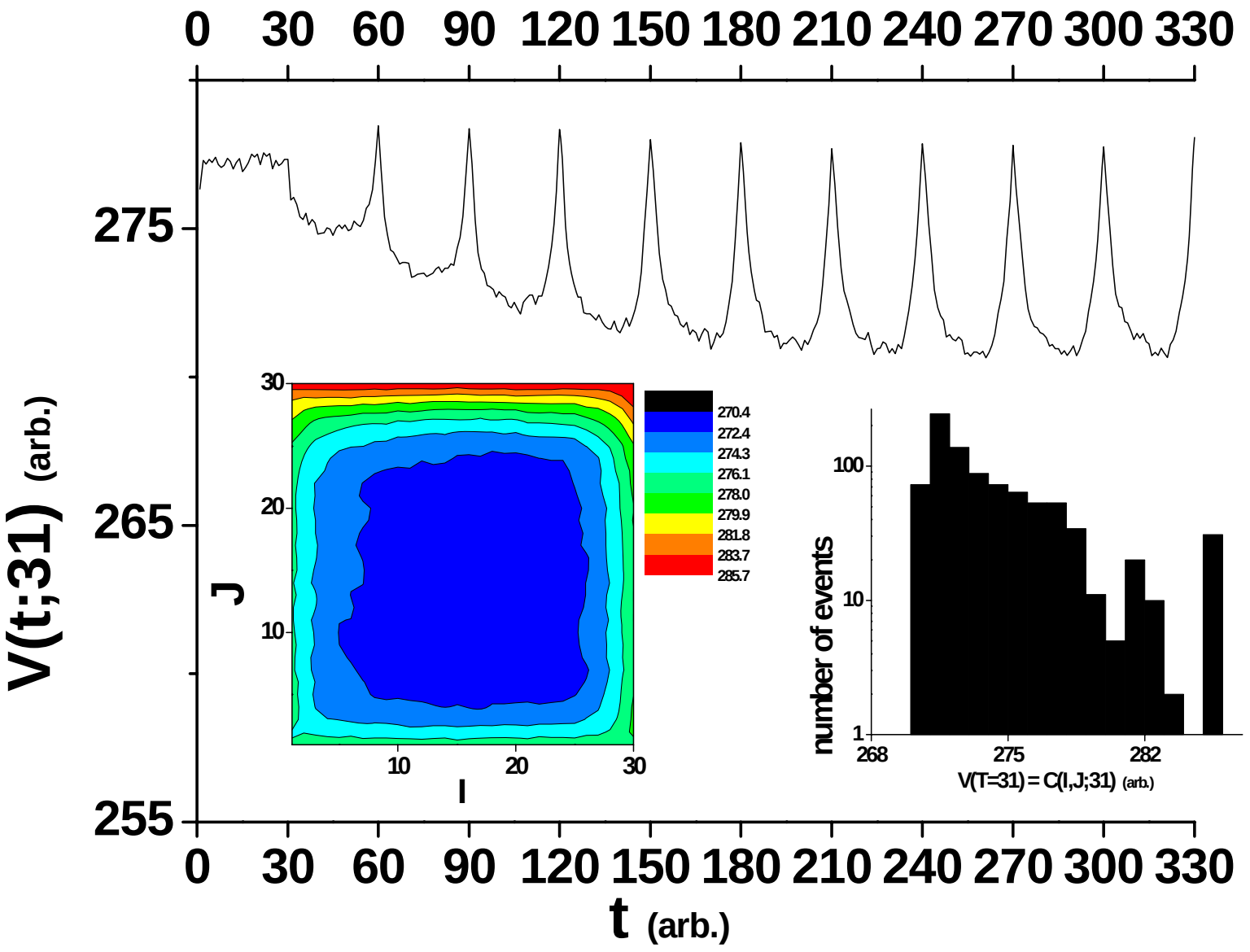

**Figure 2 (b)** A simulation for the heart beats with the parameters given in the Table I. The insets are the same as in the Figure 2 (a) but with different parameters.

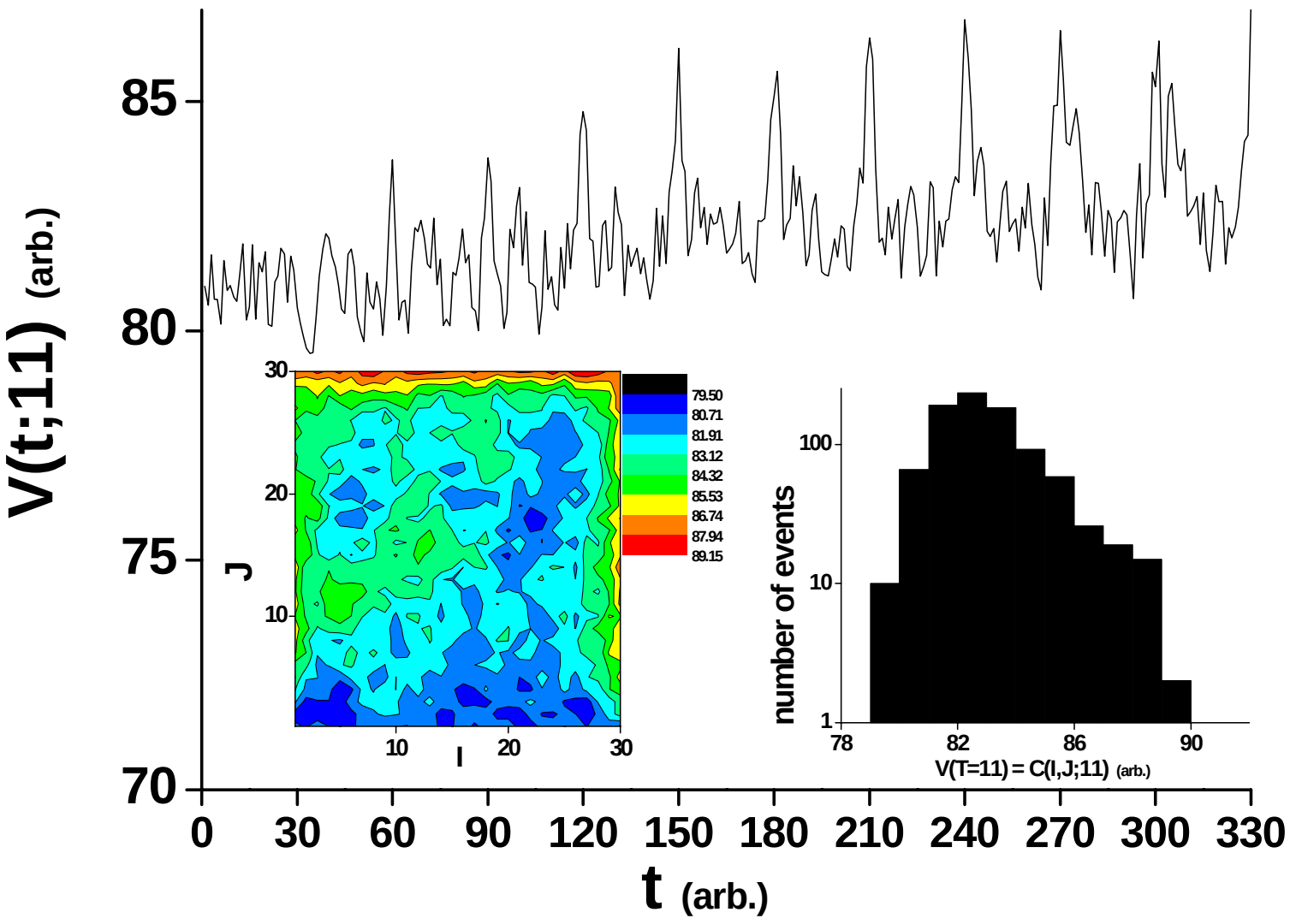

**Figure 2 (c)** A simulation for the unstable heart beats with the parameters given in the Table I. The insets are the same as in the Figure 2 (a) but with different parameters.

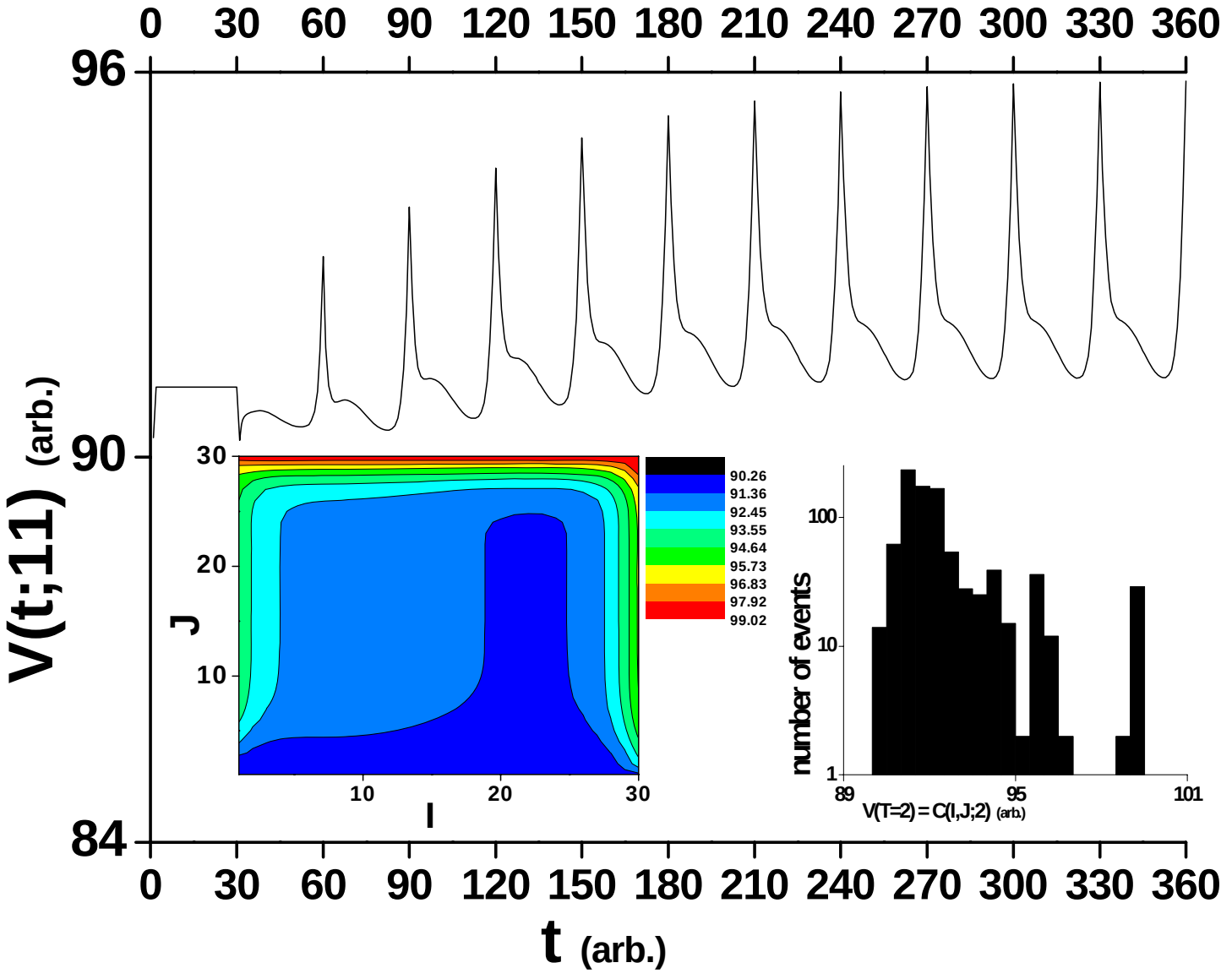

**Figure 2 (d)** Various model heart beats same as the Fig. 2 (a) but for T=11. The insets are the same as in the Figure 2 (a) but with different parameters.

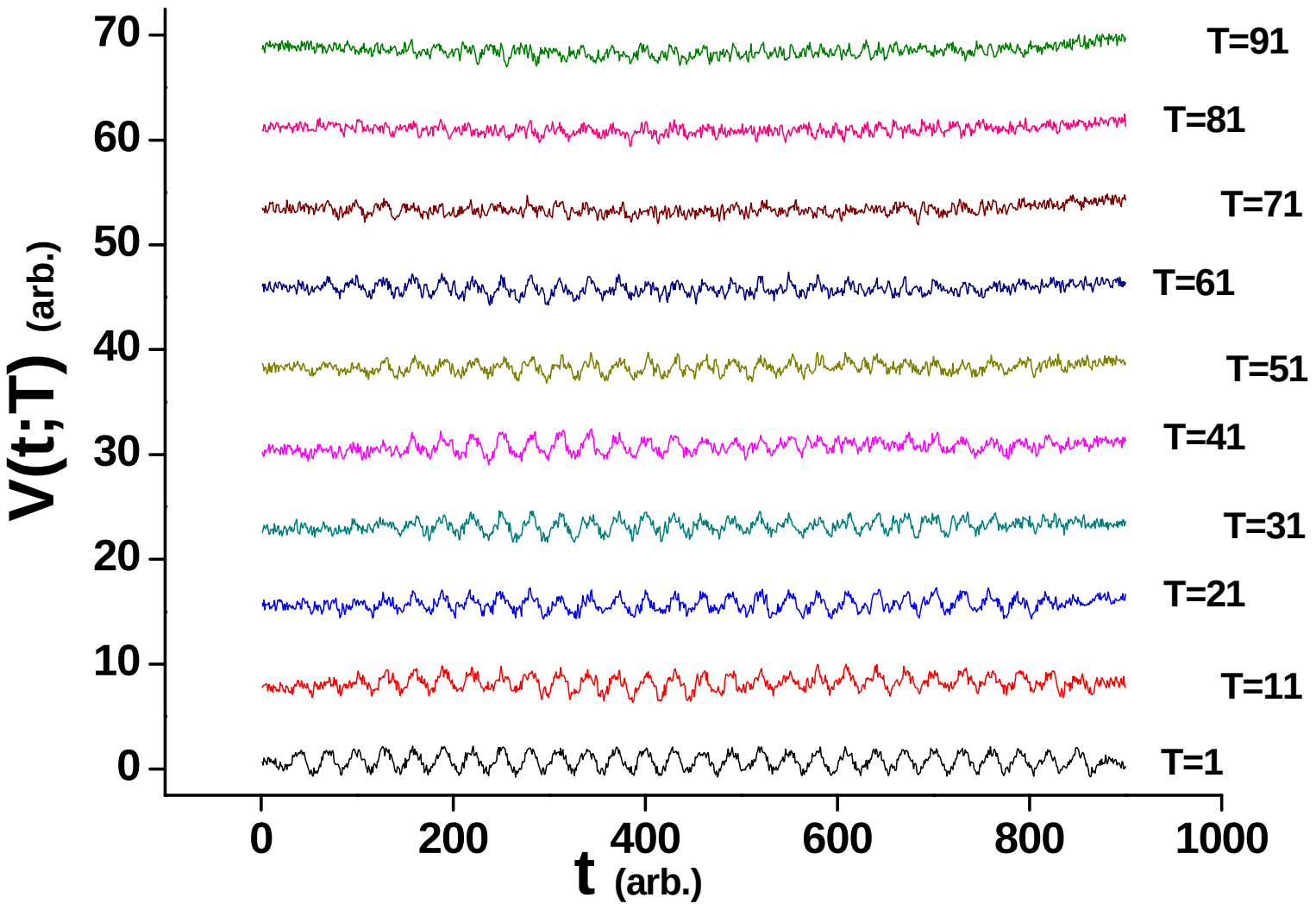

**Figure 3 (a)** The model brain signals V(t;T) for various T as designated at right (parameters are given in the Table I). The initial wave is taken as V(t;0)= – sin(2πJ/30) (=(C(I,J;T=0)).

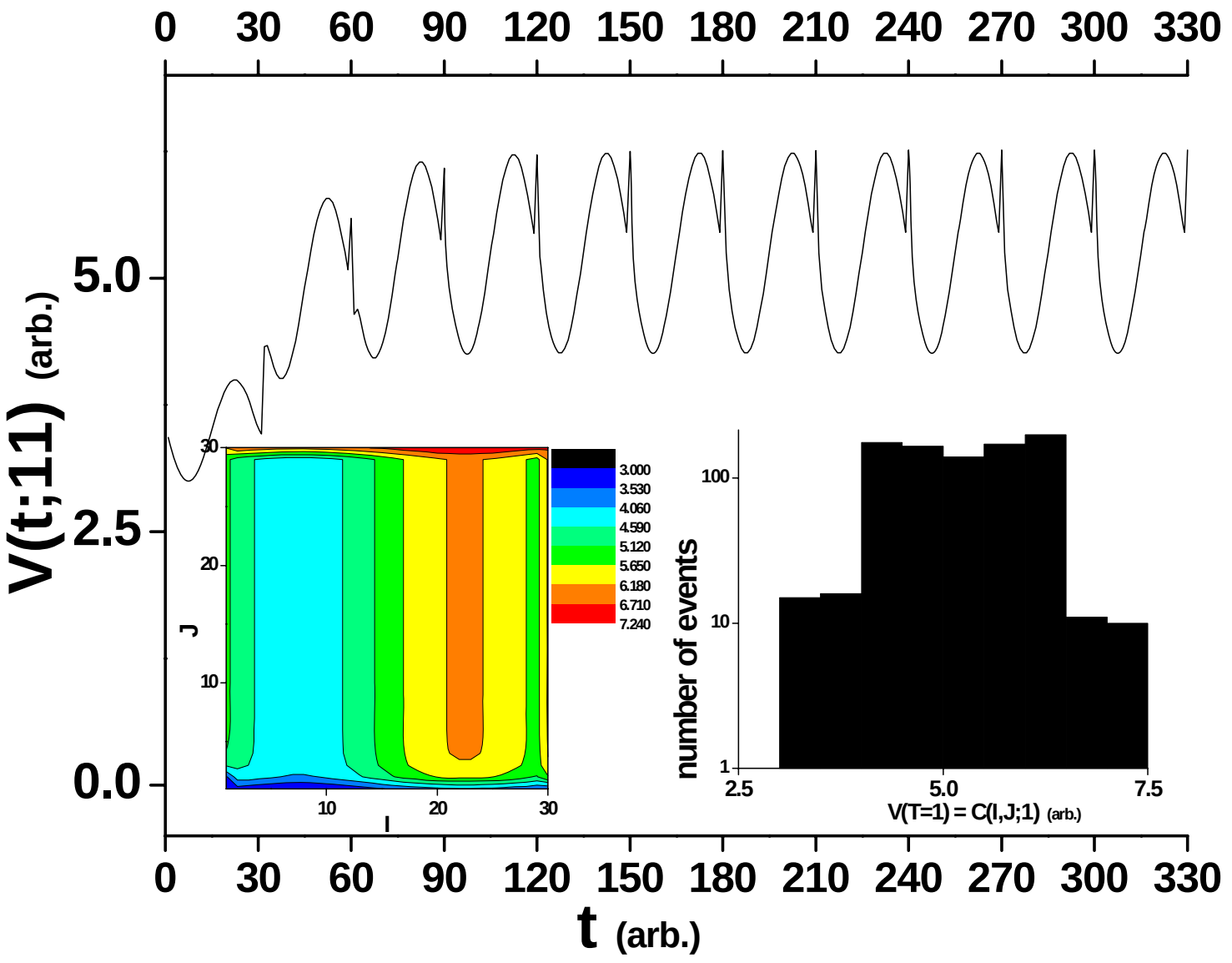

**Figure 3 (b)** The model brain signals V(t;1) with irregularity, where no random parameters are used (Table I); the designs of the pulses may be computed (not simulated) in terms of the Eq. (1) in 11 steps since T=11, where C(I,J;T=0)= – sin(2πJ/30). The insets are the same as in the Figure 2 (a) but with different parameters.